\documentclass[journal,twoside,web]{ieeecolor}
\usepackage{generic}
\usepackage{cite}
\usepackage{amsmath,amssymb,amsfonts}
\usepackage{algorithmic}
\usepackage{graphicx}
\usepackage{algorithm,algorithmic}
\usepackage{hyperref}
\hypersetup{hidelinks=true}
\usepackage{textcomp}
\usepackage{xcolor}
\usepackage{threeparttable}
\usepackage{multirow}
\usepackage{booktabs}
\usepackage{upgreek}
\usepackage{fancyhdr}

\def\BibTeX{{\rm B\kern-.05em{\sc i\kern-.025em b}\kern-.08em
    T\kern-.1667em\lower.7ex\hbox{E}\kern-.125emX}}
\begin{document}
\title{Eating Speed Measurement Using Wrist-Worn IMU Sensors Towards Free-Living Environments}
\author{
Chunzhuo Wang, \IEEEmembership{Graduate Student Member, IEEE}, T. Sunil Kumar, \IEEEmembership{Member, IEEE}, Walter De Raedt,\\ Guido Camps, Hans Hallez, \IEEEmembership{Member, IEEE}, and Bart Vanrumste, \IEEEmembership{Senior Member, IEEE} 
\vspace{-0.5cm}
\thanks{This project is funded in part by KU Leuven under grant C3/20/016, and in part by the China Scholarship Council (CSC) under grant 202007650018. \emph{(Corresponding author: Chunzhuo Wang.)}}
\thanks{This work involved human subjects in its research. Approval of all ethical and experimental procedures and protocols was granted by the Institutional Review Board (IRB) of KU Leuven under Application No. G-2021-4025-R4, and performed in line with the Helsinki Declaration.}
\thanks{Chunzhuo Wang and Bart Vanrumste are with the e-Media Research Lab, and also with the ESAT-STADIUS Division, KU Leuven, 3000 Leuven, Belgium (e-mail:chunzhuo.wang@kuleuven.be; bart.vanrumste@kuleuven.be).}
\thanks{T. Sunil Kumar is with University of Gävle, 801 76 Gävle, Sweden (e-mail: sunilkumar.telagam.setti@hig.se).}
\thanks{Walter De Raedt and Chunzhuo Wang are with the Life Science Department, IMEC, 3001 Heverlee, Belgium (e-mail: walter.deraedt@gmail.com).}
\thanks{Guido Camps is with the Division of Human Nutrition and Health, Department of Agrotechnology and Food Sciences, Wageningen University and Research, 6700EA Wageningen, and also with the OnePlanet Research Center, 6708WE Wageningen, The Netherlands (e-mail: guido.camps@wur.nl).}
\thanks{Hans Hallez is with the M-Group, DistriNet, Department of Computer Science, KU Leuven, 8200 Sint-Michiels, Belgium (e-mail: hans.hallez@kuleuven.be).} 
}
\maketitle
\thispagestyle{fancy}

\begin{abstract} 
Eating speed is an important indicator that has been widely investigated in nutritional studies. The relationship between eating speed and several intake-related problems such as obesity, diabetes, and oral health has received increased attention from researchers. However, existing studies mainly use self-reported questionnaires to obtain participants' eating speed, where they choose options from slow, medium, and fast. Such a non-quantitative method is highly subjective and coarse at the individual level. This study integrates two classical tasks in automated food intake monitoring domain: bite detection and eating episode detection, to advance eating speed measurement in near-free-living environments automatically and objectively. Specifically, a temporal convolutional network combined with a multi-head attention module (TCN-MHA) is developed to detect bites (including eating and drinking gestures) from IMU data. The predicted bite sequences are then clustered into eating episodes. Eating speed is calculated by using the time taken to finish the eating episode to divide the number of bites. To validate the proposed approach on eating speed measurement, a 7-fold cross validation is applied to the self-collected fine-annotated full-day-I (FD-I) dataset, and a holdout experiment is conducted on the full-day-II (FD-II) dataset. The two datasets are collected from 61 participants with a total duration of 513 h, which are publicly available. Experimental results show that the proposed approach achieves a mean absolute percentage error (MAPE) of 0.110 and 0.146 in the FD-I and FD-II datasets, respectively, showcasing the feasibility of automated eating speed measurement in near-free-living environments.
\end{abstract}


\section{Introduction}
\label{sec:introduction}

\IEEEPARstart{E}{ating} speed is considered as an important factor associated with body mass index (BMI), obesity and diabetes, which has been widely investigated \cite{b1,b2,b3}. Participants with faster eating speeds are considered more likely to have higher BMI, a higher risk of obesity, diabetes, and cardiovascular disease. Furthermore, deviations in eating speed are also correlated with eating disorders \cite {b4}. Currently the most popular method for investigating eating speed is through self-reported questionnaires \cite{b2, b3}. In the work of Kudo et al. \cite{b3}, participants were asked to answer questions like “How fast do you eat compared to others around same ages? (Faster, Normal, Slower).” The questionnaire based estimation is highly subjective, and there is no standard reference to define an appropriate objective eating speed. While self-reported eating speed may be sufficient at a large group level, it is an unreliable approach to assess an individual’s eating speed, particularly in the context of precision healthcare\cite{b8}. There is a call for an automated and objective approach to measure eating speed.

Recently, automated food intake monitoring has drawn lots of attention, plenty of approaches haven been proposed to detect bites \cite{b9,b11,b12,b44} during meal sessions, detect eating episodes in free-living scenarios\cite{b13,b14} using various sensors (e.g., inertial, camera, microphone, proximity). However, to date, there has been no research that focus on automated eating speed detection in (near-)free-living environments. 

In this study, we use the term \emph{bite} to refer to eating and drinking gestures. The definitions of eating and drinking gestures are consistent with the work in \cite{b26}. Specifically, they are defined as the action of raising the hand to the mouth with cutlery until the hand is moved away from the mouth. The definition of objective eating speed is the number of bites divided by the time taken to finish the eating episode (bites/min), which has been utilized in \cite {b1,b8, b6}. Based on this definition, a straightforward approach for automated eating speed estimation in full-day scenario is to combine the bite detection (to count the number of bites) and eating episode localization (to obtain the time duration of the consumed eating episode). However, the reason that hinder automated eating speed estimation is two-fold: Firstly, existing bite detection approaches only focus on in-meal scenarios, it is challenging to detect bites in (near-)free-living environments. Secondly, current eating episode localization approaches cannot precisely segment the boundary of detected eating episode, whereas eating speed detection requires accurate boundary segmentation. 

In this study, we combine two classical tasks in food intake monitoring, i.e., bite detection and episode detection, to facilitate the automated eating speed estimation. The main contributions of this research can be summarized as follows:
\begin{itemize}
\item A complete framework for eating speed measurement in near-free-living environments is proposed: A sequence-to-sequence (seq2seq) temporal convolutional network combined with a multi-head attention (TCN-MHA) model is designed to process inertial measurement unit (IMU) data for detecting food intake gestures and segmenting the time interval of intake gestures. The obtained bite sequences are clustered into eating episodes to calculate eating speed. 

\item An intensive comparison between our approach and existing works has been implemented. The proposed eating speed measurement method is validated in two studies: 7-fold cross validation on the well-annotated full-day-I (FD-I) dataset, and a holdout validation on the coarsely annotated full-day-II (FD-II) dataset. Additionally, the public OREBA and Clemson are utilized to further analyze the performance of proposed approach. To our best knowledge, this is the first work to automatically estimate eating speed in near-free-living environments.

\item We make two datasets collected in this study publicly available\footnote{https://rdr.kuleuven.be/dataset.xhtml?persistentId=doi:10.48804/CN8VBB}. Specifically, the FD-I dataset contains IMU data collected from 34 participants in near-free-living environments with fine annotation. This is the first full-day IMU dataset that contains eating and drinking gesture annotations not only during meal sessions, but also outside of meal sessions. The FD-II dataset serves as a holdout dataset, containing IMU data from 27 participants in free-living environments.
\end{itemize}

\section{Related work}
Eating speed detection relies on two essential tasks: bite detection to count the number of bites and eating episode detection to predict the duration. In this section, we firstly introduce existing approaches for in-meal bite detection; then, we discuss eating episode detection approaches. Thirdly, we present a few studies measuring eating speed objectively. Finally, deep learning for time-series signal are introduced.

\vspace{-0.25cm}
\subsection{Bite Detection}
Bite detection has been widely investigated using various sensors. Cameras have been used to detect bites \cite{b45,b46} and food types \cite{b11}. Acoustic sensors can be used for chewing sounds detection\cite{b12}. Mertes et al. \cite{b47} developed a strain gauge-based smart plate to detect bites based on the weight change of food. In our recent work \cite{b26}, a novel radar-based system was validated using our public Eat-Radar dataset. The photoplethysmography (PPG) sensor \cite{b49} and electromyography (EMG) sensor \cite{b50} have also been explored for bite detection. Apart from these sensors, to date, the wrist-worn IMU sensor is a popular choice for bite detection because of its least burdensome and most acceptable. Dong et al. \cite{b15} developed a rule-based approach to detect bite using the rotation velocity of wrist. Shen et al. \cite{b16} further evaluated Dong's approach on Clemson dataset. Kyritsis et al. \cite{b9} proposed an end-to-end based approach using a convolutional neural network and long-short-term-memory network (CNN-LSTM) model to detect bite on FIC dataset. Rouast et al. \cite{b17} further developed single-stage ResNet based CNN-LSTM architecture for bite detection on the OREBA dataset. Wei et al. \cite{b28} developed an energy-efficient approach integrating an optimized multicenter classifier (O-MCC) to detect intake gestures with low inference time.

The aforementioned approaches have shown promising performances. However, it should be noted that these approaches only focus on bite detection within meal sessions (10-20 min). A more challenging scenario, bite detection in (near-)free-living environments ($\geq{6}$ h), has yet to be broadly investigated. There are several obstacles impeding the detection. Firstly, it is troublesome to obtain bite-level label outside meal sessions. Secondly, in the scale of full-day duration, bites are extremely sparse, leading a highly imbalanced dataset compared to in-meal datasets, making bite detection more challenging.

\begin{figure*}[t]
      \centering
      \includegraphics[scale=0.66]{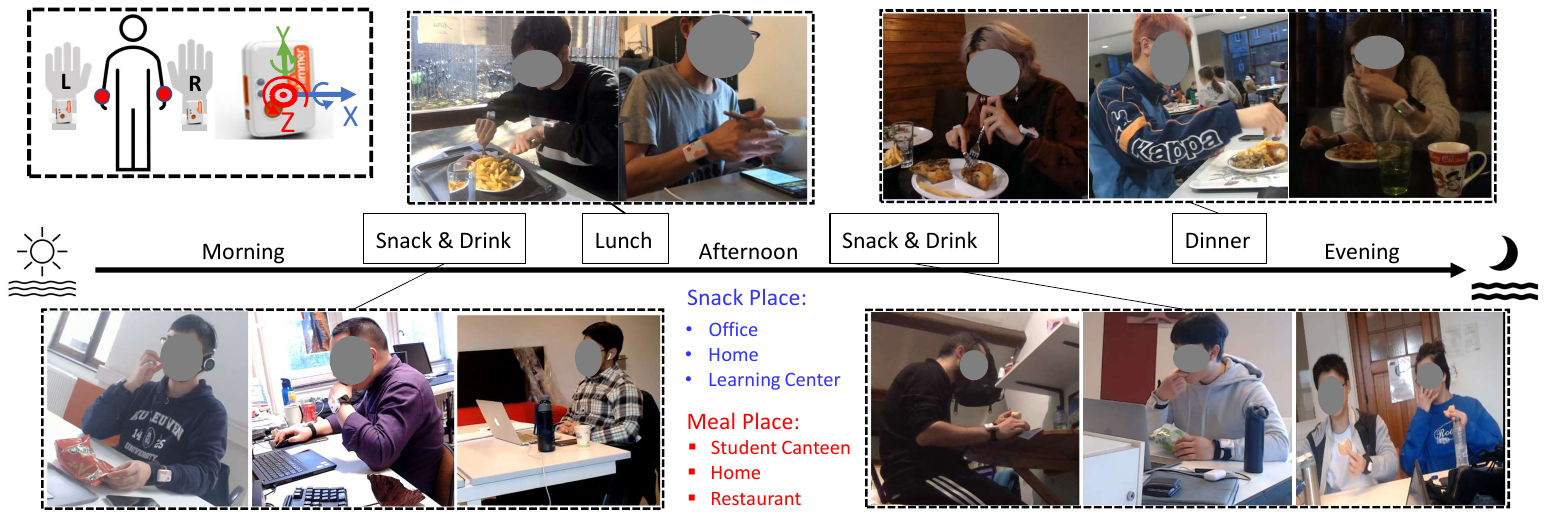}      
      \vspace{-0.2cm}
      \caption{Examples of wrist-worn IMU data collection. Two IMU sensors were mounted on both hands. Participants completed their daily activities without restriction. This figure only shows the food intake related scenes, their other daily activities, such as studying, walking, talking, running, cooking, were recorded by IMU sensors as well.}  
      \vspace{-0.3cm}
      \label{data_collection}
\end{figure*}

\begin{figure}[t]
      \centering
      \includegraphics[scale=0.66]{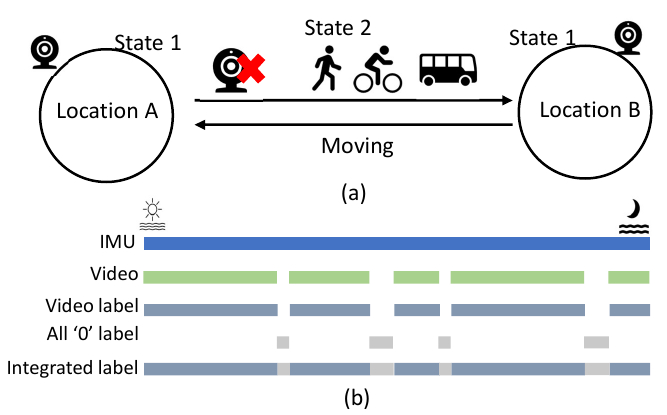}      
      \vspace{-0.2cm}
      \caption{(a) the data collection protocol; (b) the annotation method, the video label is for state 1, and the all '0' label is for state 2.}  
      \vspace{-0.3cm}
      \label{protocol_syn}
\end{figure}

\vspace{-0.25cm}
\subsection{Eating Episodes Detection}
Eating episodes detection is another popular research topic in food intake monitoring. Such a system mainly focus on the detection of eating episodes under free-living environments. A standard process pipeline involves cutting the full-day data into minute-level segments, and machine learning is used to predict if each segment belongs to eating episode or not. Sharma et al. \cite{b13} collected a dataset including 354 days of wrist-worn IMU data from 351 subjects. Doulah et al. \cite{b14} developed the AIM-2 system, a pair of eyeglasses mounted with a camera, and a 3-axis accelerometer to detect eating episodes.

Unlike the above methods that directly predict eating episodes, another routine is first to detect basic elements of eating episode such as chewing, swallowing, and hand-to-mouth events, and then combine them together as an eating episode using various merging techniques. Bedri et al. \cite{b18} proposed the FitByte eyeglass-based system to detect intake events and eating episodes. They detected the episodes by merging any detected intakes that are within 5 minutes from each other. Zhang et al. \cite{b19} developed Necksense system to detect chewing sequences and cluster detected sequences into eating episodes using a density-based spatial clustering of applications with noise (DBSCAN) algorithm \cite{b42}. Kyritsis et al. \cite{b9} first detected bites on FreeFIC dataset, then applied a Gaussian filter on the bite sequence to generate meal regions. Noted that the model used to detect bites is trained on in-meal FIC dataset, there is no bite-level label in FreeFIC dataset, so the bite detection performance is unclear on FreeFIC dataset.

\vspace{-0.25cm}
\subsection{Eating Speed Estimation}
Several studies have been found to objectively (but not automated) measure eating speed. Woodward et al. \cite{b8} provided a 550 g meal to each participant and used a stopwatch to record the meal time in lab environments. Scisco et al. \cite{b48} explored the utility of the Bite Counter device during meal sessions, and further utilized the device in free-living environments to estimate eating speed \cite{b49}. Notably, in  \cite{b49}, users were required to press a button to start and stop the counter manually, and the accuracy of bite detection was unclear due to the absence of ground truth bite information. Alshurafa et al. \cite{b6} measured eating speed by using wearable fish-eye camera in free-living environments. A trained annotator manually counted the number of intake gestures, selected the boundaries of meal sessions by viewing the video. The eating speed was then calculated by using meal duration to divide the number of bites.  

\vspace{-0.25cm}
\subsection{Temporal Sequence Models}
The commonly used models in intake gesture detection are CNN with recurrent neural networks (CNN-RNNs). The CNN is typically used to extract features from time-series data, and the feature sequences are then fed into RNNs to process temporal dependencies. Although RNNs can effectively process time-series data, they struggle to memorize long-term interdependencies due to the gradient vanishing problem. Two solutions have been proposed in the literature: the temporal convolutional network (TCN) and self-attention.
\subsubsection{TCN}
Lea et al. \cite{b21} proposed the TCN by utilizing dilated convolution and residual connection. Stacking a series of convolution layers with different dilation factors enables the model to incorporate short-term and long-term dependencies. Additionally, dilated convolutional layers increase the length of receptive fields without substantially increasing the number of parameters. Due to its superiority, studies have been conducted to further evolve the architecture (e.g. MS-TCN \cite{b23}, MS-GCN \cite{b24}) and to exploit it into various scenarios.  
\subsubsection{Self-Attention}
The self-attention module was proposed by Vaswani et al. \cite{b25} to compose the transformer architecture in natural language processing (NLP) domain. Motivated by its success, several approaches have been proposed by integrating the module into CNN and RNN architectures to further improve the capability of time-series signal modeling \cite{b40,b41}.

\section{Methods}

\subsection{Sensors}
As this experiment aims to record data in (near-)free-living environments, food intake events can happen with both hands, thus, two shimmer3 IMU wristbands\footnote{https://shimmersensing.com/product/shimmer3-imu-unit/} were mounted on both hands of participants. The battery duration of shimmer3 IMU is 24 h, which satisfies the requirement of this experiment. The sampling frequency was set to 64 Hz. The 3-axis accelerometer and 3-axis gyroscope units were activated to generate 6 channels IMU data. The data were stored in SD card of shimmer, and can be downloaded into laptop via Consensys software\footnote{https://shimmersensing.com/product/consensyspro-software/} for further data processing. A camera was used to record the experiment for annotation. The sensor deployment and data collection example are shown in Fig. \ref{data_collection}.

\vspace{-0.25cm}
\subsection{Full-Day Data Collection}
This research was approved by the ethical committee of KU Leuven with project number: G-2021-4025-R4. Informed consent was obtained from all participants. Two datasets were collected for this study, namely the well-annotated full-day-I (FD-I) dataset, and holdout full-day-II (FD-II) dataset. There is no participants overlap among the two datasets. The statistics of the datasets are shown in Table \ref{fd_data}.

\subsubsection{FD-I Dataset}
\begin{figure}[t]
  \centering
  \includegraphics[scale=0.25]{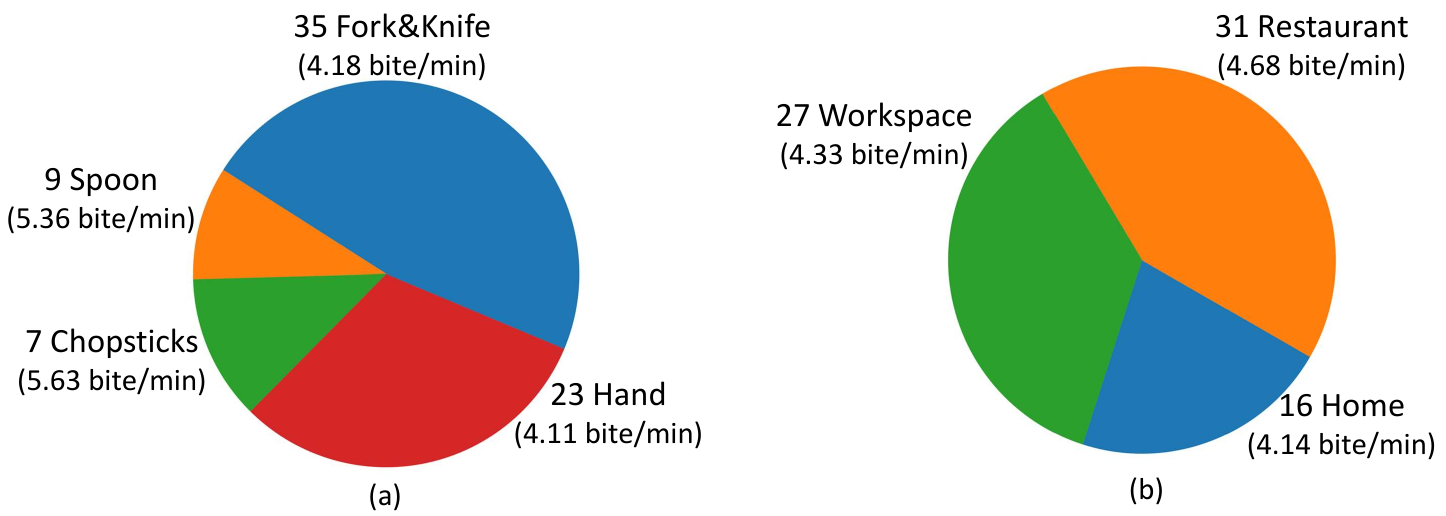}
  \vspace{-0.2cm}
  \caption{The pie chart for eating styles (a) and eating locations (b). The eating speeds represent the average speed of each category.}
  \vspace{-0.2cm}
  \label{pie_stats}
\end{figure}

{
The FD-I dataset contains 34 days of IMU data from 34 participants (6 of them are from our previous study on drinking activity detection \cite{b43}). On the data collection day, our research assistants met the participant, instructed the participant to wear IMU wristbands. Participants were free to engage in their normal daily activities. Our research assistants were responsible for the recording when participants change their locations to ensure that all eating and drinking gestures were captured. Specifically, participants were accompanied by one of our research assistants. To clarify the data collection protocol, we have formulated the participants' daily activities into two states, as illustrated in Fig. \ref{protocol_syn}(a). In state 1, the participant was engaged in activities (e.g., studying, working, and taking meals) at a specific location (Location A). In state 2, the participant was transiting from Location A to Location B by walking, biking, or taking a bus. During state 1, the camera recorded continuously to ensure all intake gestures were captured, especially outside of meal sessions. However, during state 2, we did not record under the assumption that no intake gestures occur during the transition. Specifically, recording started upon the participants' arrival at Location A and stopped upon their departure. No recording took place during their transition from Location A to Location B. A new recording started upon their arrival at Location B. The assistant mainly used a Logitech camera connected to a laptop to record video during state 1. There were no restrictions placed on participants' activities or locations during data collection. Using a small photography tripod, the orientation and height of the camera could be easily adjusted to facilitate the deployment. At the beginning of every video recording, the timestamp was presented to enable the synchronization of IMU and video data. At least one meal was collected from each participant, and the minimum data collection duration was 6 h. Both solitary and social eating scenarios were included in the dataset. Participants received a restaurant voucher (20 euro) as experiment compensation after the data collection. A total of 251.70 h two-hand IMU data were collected, which contains 74 eating episodes, with 4,568 eating and 1,100 drinking gestures. The dataset contains four eating styles including fork \& knife, chopsticks, spoon, and hand, as shown in Fig. \ref{pie_stats}(a). Eating locations in this dataset include participant's home (apartment, student residence), restaurant, and workspace (library, university learning center, and campus rest areas), as shown in Fig. \ref{pie_stats}(b). 

According to the data collection protocol, we recognize that the experimental conditions differ from controlled meal sessions, which are characterized by a single location and short duration. Our approach involves extended monitoring periods and multiple locations within participants' daily routines. However, these conditions do not fully represent free-living environments as a research assistant consistently accompanies the participants during data collection to ensure the capture of all intake gestures. This presence may introduce additional interference in participants' daily activities. Therefore, we employ the term \emph{near-free-living environments} to describe our experimental conditions.

}

\subsubsection{FD-II Dataset}
{
The FD-II dataset contains 27 days of IMU data from 27 participants. The experiment protocol was the same as the FD-I dataset. However, in this dataset, only meal sessions were recorded by cameras (Some videos were collected by participants' own smartphones). All other eating/drinking gestures outside of meal sessions were not recorded. Therefore, the ground truth information only contains the bite information during meals and the meal boundaries. The FD-II dataset is considered as a holdout dataset, which contains 52 meals with 2,722 eating gestures (including four eating styles) over a total duration of 261.68 h.
}

\vspace{-0.15cm}
\subsection{In-Meal Datasets}

 The FD-I and FD-II datasets are highly unbalanced, with the target classes being the minority. To further include more target data, we included two datasets containing IMU data collected in meal sessions, specially, the self-collected meal-only (MO) dataset and the external OREBA dataset \cite{b27} as part of the training set. Additionally, the OREBA and Clemson \cite{b16} datasets were included for comparison analysis.

\subsubsection{Meal-Only Dataset}
{
The MO dataset contains 46 meal sessions from 46 participants, including 2,894 eating and 763 drinking gestures. It should be noted that part of this dataset was collected together with our Eat-Radar project \cite{b26} and there is no participant overlap between MO and FD datasets. 
}
\subsubsection{Public OREBA Dataset}
{
The OREBA dataset \cite{b27} contains 100 meal sessions data from 100 participants, with 4,496 eating and 406 drinking gestures. The data were collected from both hands using two IMU wristbands. We downsampled the data from 64 Hz to 16 Hz. It should be noted that the coordinate system (direction of $x$, $y$, and $z$ axis) of the sensor used in OREBA is the same as ours, allowing us to integrate this data into the training set.
}

\subsubsection{Public Clemson Dataset}
{
The Clemson dataset \cite{b16} contains 488 meal sessions across 264 participants, with a total of 18,430 eating and 2,152 drinking gestures. The data were collected by placing a wristband IMU on the dominant hand and the sampling rate of data is 15 Hz.}

\begin{table}[t]
\caption{Full Day Datasets Statistics}
\label{fd_data}
\begin{center}
\scalebox{0.7}{
\begin{tabular}{l|cc}
\toprule
\hline
 Parameter & FD-I & FD-II \\
\midrule
 \# Participants  & 34 & 27  \\
 \# Days          & 34 & 27  \\
 \# Eating episodes   & 74 & 52  \\
 \# Eating gestures & 4,568 & 2,722  \\
 \# Drinking gestures & 1,100 & -  \\
 Mean day duration (h) &7.40$\pm$2.13  &9.69$\pm$3.77  \\
 Duration ratio of other : eating : drinking  &  142.52 : 2.51 : 1&  116.43 : 1 : -\\
\hline
\bottomrule
\end{tabular}}
\end{center}
\vspace{-0.3cm}
\end{table}

\begin{figure}[t]
  \centering
  \includegraphics[scale=0.45]{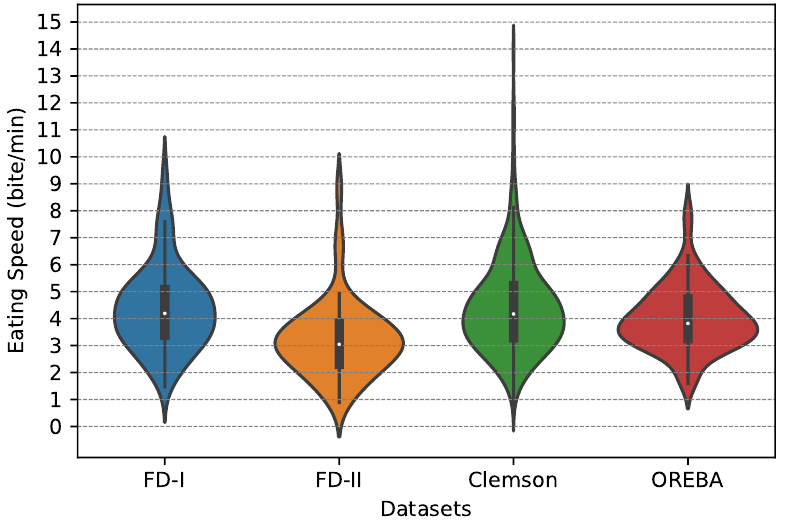}
  \vspace{-0.2cm}
  \caption{The violin plot of ground truth eating speed distribution among different datasets.}
  \vspace{-0.2cm}
  \label{speed_violin}
\end{figure}

\vspace{-0.15cm}
\subsection{Annotation}
\subsubsection{Bite Annotation}
Video recordings were viewed to annotate bites via ELAN \cite{b29}. The data were labeled into 3 classes, eating gesture, drinking gesture, and others. Three trained annotators labeled the datasets, with each annotator assigned to a specific portion of the dataset. The first author rechecked the annotation and made corrections when necessary. All annotators followed the same annotation instructions. To synchronize the video and IMU data, the IMU data were divided into segments using the timestamps recorded at the start of each video, as described in Section III-B-1. For IMU segments corresponding with video recordings (state 1), the timelines were aligned by matching the IMU timestamps with those at the beginning of each video. The ELAN tool was used to import both the video and IMU data to verify the alignment and manually adjust it if there was a time shift (normally 1-2 s). Timing shift was identified by observing whether the IMU waveform response to the participants' hand movements, transitioning from stillness to motion and vice versa, was delayed or advanced. For IMU segments without video recordings (state 2), labels of '0' were assigned directly. These labeled segments were then concatenated together, as shown in Fig. \ref{protocol_syn}(b).
\subsubsection{Eating Episodes Annotation}
The first eating gesture in an eating episode signifies the beginning boundary of the eating episode, and the last eating gesture in an eating episode is considered the ending boundary of the episode. In free-living environments, it is normal for people to eat snacks outside meal sessions. However, snack eating exhibits high variability compared to meal eating. Some snack eating has frequent bites in very short duration, while other snack eating occurs over a longer period with a very low frequency (only one bite in several minutes). This variability makes the detection of snack session difficult; therefore, we focus on episodes that last at least 3 min. Snacking sessions with duration less than 3 min were neglected in this step. According to our definition, there are 457 episodes in Clemson and 101 episodes in OREBA. 
\subsubsection{Ground Truth Eating Speed}
The ground truth eating speed is obtained by computing the ratio of the number of bites in the eating episode and the duration of the episode. The unit of eating speed is bites per minute (bite/min). Fig. \ref{speed_violin} shows the distribution of eating speed in different datasets.

\begin{figure}[t]
  \centering
  \includegraphics[scale=0.41]{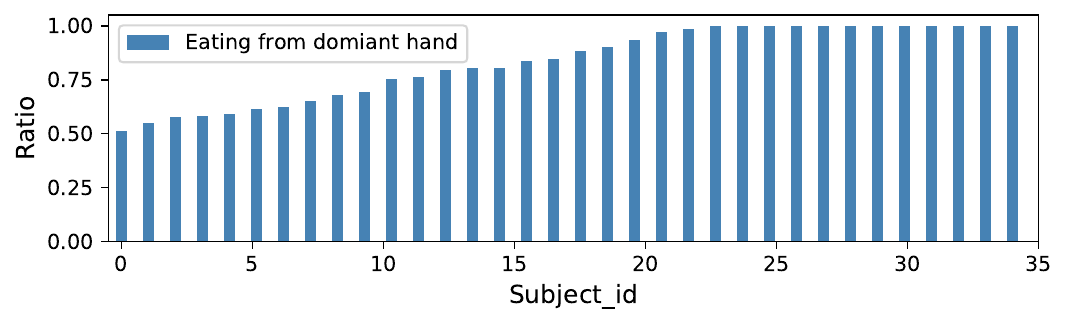}
  \vspace{-0.2cm}
  \caption{The quantity ratio of eating gestures using only dominant hand on FD-I dataset. A ratio with 0.5 means half number of eating gestures are from the dominant hand, the others are from the non-dominant hand.}
  \vspace{-0.3cm}
  \label{dominant_hand}
\end{figure}

\begin{figure}[t]
  \centering
  \includegraphics[scale=0.40]{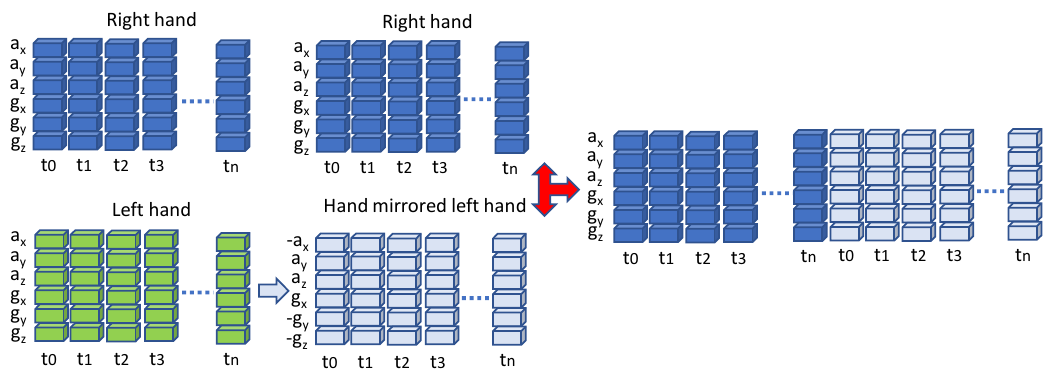}
  \vspace{-0.2cm}
  \caption{Visualization of the two hand combination process. $t_n$ represents the end time point of the eating episode.} 
  \vspace{-0.3cm}
  \label{two-hand}
\end{figure}

\vspace{-0.25cm}
\subsection{Data Preprocessing}
In daily life, both hands can be used for eating food, as illustrated in Fig. \ref{dominant_hand}. Considering the IMU waveform for eating with left and right hand differs, we applied the two-hand combination method that combines hand mirroring and temporal concatenation to process two-hand IMU data, which was validated in our previous study \cite{b5}. The hand mirroring method has been applied in multiple IMU-based bite detection studies when the participant is left hand dominant, which involves flipping the direction of $a_x$ (in accelerometer), $g_y$ and $g_z$ (in gyroscope). The hand-mirrored left hand IMU data were than concatenated after right hand data. Hence, the preprocessed data have 6 channels, and the data length of each recording is doubled, as shown in Fig. \ref{two-hand}. Meanwhile, to reduce the computation cost, the data were downsampled to 16 Hz.

\begin{figure*}[t]
      \centering
      \includegraphics[scale=0.70]{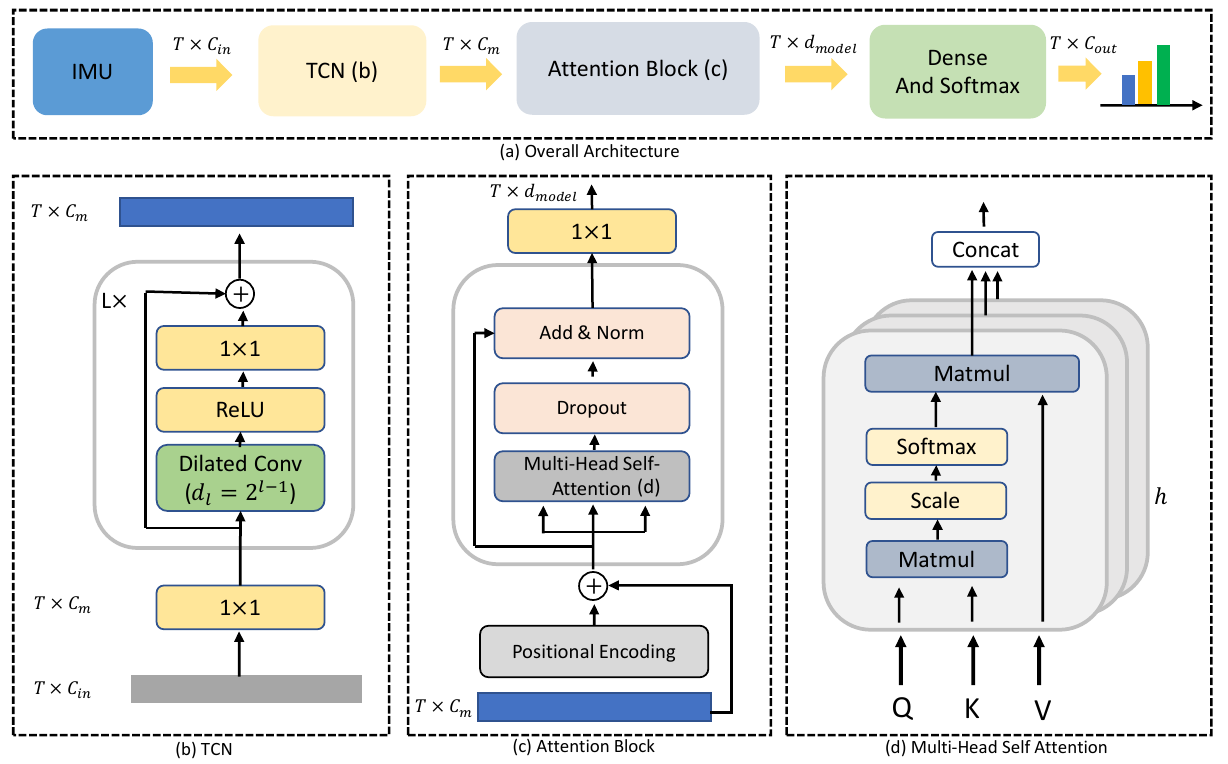} 
      \vspace{-0.2cm}
      \caption{The architecture of the proposed TCN-MHA model. Figure (a) shows the overall architecture of the model. The processed IMU data are first fed into the TCN module, followed by further processing of the TCN module's output through the Attention block. Finally, the FCN block is applied to process the output of the attention block to generate predictions. Figure (b) represents the architecture of TCN module. The architecture of attention block is illustrated in Figure (c). Figure (d) explains the mechanism of the multi-head attention module. $d_l$ signifies the dilated factor, where $l$ denotes the layer order, $L$ represents the total number of TCN layers, $h$ represents the number of attention heads.} 
      \vspace{-0.2cm}
      \label{bite-model}
\end{figure*} 

\vspace{-0.25cm}
\subsection{Deep Learning Model}
To explore the potential of utilizing both TCN and attention, the proposed TCN-MHA architecture is comprised of three parts: the TCN module using dilated convolution layer to process multi-scale temporal patterns, the multi-head attention (MHA) module to further focus on representative temporal features to improve the performance, and the fully connected network (FCN) to generate predictions, as shown in Fig. \ref{bite-model}. 

\subsubsection{TCN Module}
The TCN distinguishes classical CNN by using dilated convolutions \cite{b21}. A series of dilated convolution layers are stacked together to compose the TCN module. Each layer involves conv1d with a dilation factor $d_l =2^{l-1}$ $(1\leq{l}\leq{L})$, where $l$, $L$ are the number of the current layer and total layers, respectively. A residual connection is also applied in each layer to combine the input features of current layer and the processed features. The stacked dilated layers increased the receptive field without substantially increasing the number of parameters, allowing the TCN to effectively capture long-term temporal dependencies. The structure is shown in Fig. \ref{bite-model} (a). Given the input sequences $X\in\mathbb{R}^{T\times{C_{in}}}$, a $1\times{1}$ convolution layer is first applied to adjust the dimension to $T\times{C_m}$, where $C_{in}$ is the input dimension, and $C_m$ equals to the number of kernels in each dilated convolution layer (the number of kernels of each conv layer is the same). After $L$ dilated conv layers, the output of the TCN module is $X_{1}\in\mathbb{R}^{T\times{C_m}}$.
\subsubsection{MHA Module}
The MHA module is an essential component of the Transformer \cite{b25}, showcasing considerable efficacy in capturing relationships within time sequence signals. Before being fed into the attention module, the positional encoding is added to the input feature map. The attention mechanism is considered as a mapping among the query ($Q=X_{1}W_{Q}$), key ($K=X_{1}W_{K}$), and value ($V=X_{1}W_{V}$), where $X_{1}\in\mathbb{R}^{T\times{d_{in}}}$ signifies the input of the module ($d_{in}=C_m$), $W_{Q}$,$W_{K}$, and $W_{V}\in\mathbb{R}^{d_{in}\times{d_{model}}}$ are weight matrices used to transform the input data into Q, K, and V, respectively, with $d_{in}$ representing the dimension of the output from the TCN module, and $d_{model}$ denoting the dimension of the attention module. The formula of the attention calculation is presented as follows:
\begin{equation}Att (Q,K,V) ={\rm Softmax}\left(\frac{QK^\top}{\sqrt{d_{model}}}\right)V\label{eq1}\end{equation}

The aforementioned computation involves a singular attention graph; however, multi-head attention employs multiple attention graphs to further learn attention maps across diverse aspects. The parameter $h$ denotes the number of heads, signifying the quantity of attention graphs in use.
\begin{equation}
\begin{split}
 MHA (Q,K,V) ={\rm Concat}(SA_{1},\ldots,SA_{h})W^{O}\\
{\rm where}\ SA_{i}=Att(QW_{i}^{Q},KW_{i}^{K},VW_{i}^{V})
\end{split}
\label{eq2}\end{equation}
where $W^{O}\in\mathbb{R}^{h\cdot d_{h}\times{d_{model}}}$, $W_{i}^{Q}$, $W_{i}^{K}$ and $W_{i}^{V}\in\mathbb{R}^{d_{model}\times{d_{h}}}$, and $d_h$ denotes the dimension of each individual head in the attention block, here $d_h=d_{model}/h$.

\subsubsection{FCN Module}
The output of the MHA module is fed into the final FCN classifier block containing 2 linear layers, yielding $Y\in\mathbb{R}^{T\times{C}_{out}}$, where $C_{out}$ is the number of classes. 

\subsubsection{Implementation Details}
The model's loss function comprises both classification and smoothing loss. Specifically, the classification loss is a cross-entropy loss, while the smoothing loss involves a truncated mean squared error (MSE). Details on the integrated loss function can be referenced in \cite{b23}. Following each dilated layer in the TCN module, a dropout rate of 30\% is applied. Based on experiments, the TCN module is constructed with a total of 9 layers, with each layer comprising 64 kernels. For the MHA module, 8 attention heads are employed, with each head characterized by a dimension of 16, resulting in a model dimension of 128 ($d_{model}$=128). The first layer of the FCN has 64 neurons with ReLU activation, the last layer contains 3 neurons with Softmax activation. For model training, an Adam optimizer is utilized with a learning rate set to 0.0005. It is important to note that the model exhibits a temporal lag in its predictions due to its non-causal nature. This temporal delay can be quantified by dividing half of the receptive field by the sampling frequency (0.5$\times$1023/16=32s). The window length of input data is set to 60 s accordingly. All experiments were carried out on two NVIDIA P100 GPUs provided by Vlaams Supercomputer Centrum (VSC) \footnote{See https://www.vscentrum.be/}.

\vspace{-0.15cm}
\subsection{Bite Detection}
The argmax is applied on the probability sequence generated by the TCN-MHA model to yield point-wise predictions. In order to mitigate the impact of noise in the predictions, adjacent bites intervals with identical values that are within 0.5 s are consolidated. After that, bites with duration less than 1 s are excluded. Subsequently, we obtain a set of detected bites $\mathcal{B}=\left\{b_1, b_2, ...b_N\right\}$, where each $b_i$ corresponds to the interval $[t_{i}^{l},t_{i}^{r}]$ ($1\leq{i}\leq{N}$) representing the left and right temporal boundaries of the $i$-th bite, $N$ denotes the total number of detected bites. It should be noted that the value of the interval distinguishes the type of bite (Eating: 1, Drinking: 2).

\begin{figure}[t]
      \centering
      \includegraphics[scale=0.33]{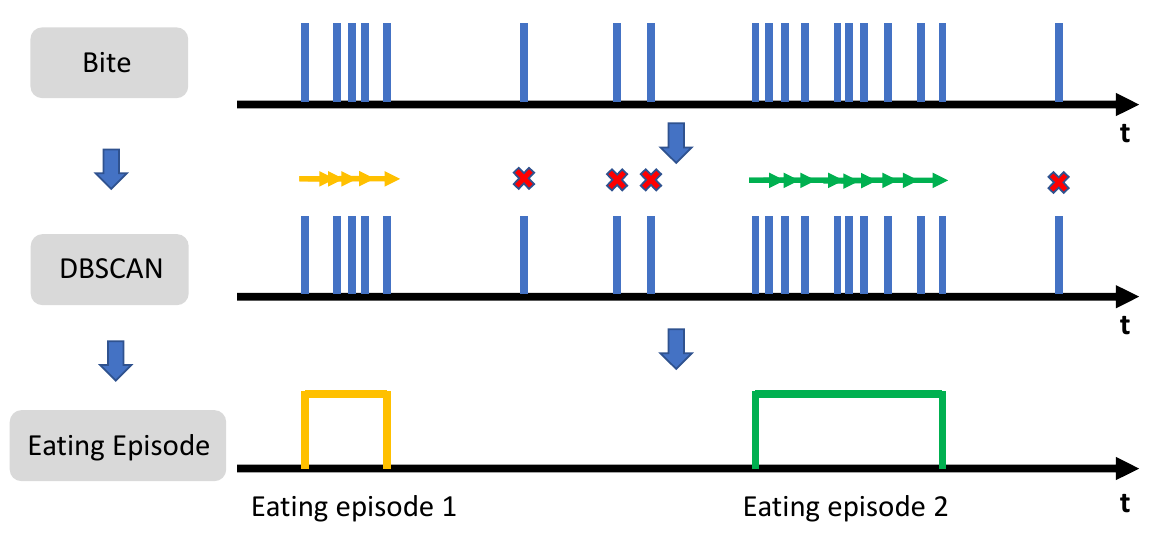}      
      \vspace{-0.2cm}
      \caption{The DBSCAN clustering example for eating episode detection. In the DBSCAN clustering step, arrows with different colors (yellow or green) represent different clusters.} 
      \vspace{-0.4cm}
      \label{dbscan}
\end{figure} 

\vspace{-0.25cm}
\subsection{Eating Episode Detection}
In bite detection step, to achieve data uniformity, the data from the left hand is hand mirrored and then temporally concatenated after right hand data. Prior to meal session detection, the output data is divided into to two sub-sequences (left hand and right hand), then an OR operation is applied to integrate bites from both hands. As eating episodes mainly involves eating gestures, all detected drinking gestures are removed in this step. The predicted bite sequence is then clustered by 1D-DBSCAN \cite{b42} to compose eating episodes. The DBSCAN identifies clusters based on the density of points. The function \emph{sklearn.cluster.DBSCAN} is employed with an epsilon parameter set to 3 min and a minimum samples parameter set to 5. In our case, the distance between two bites is the temporal proximity between the bites. Subsequently, sparse bites are filtered as noise, while bites sequence with high density are clustered together to compose eating episodes, as shown in Fig. \ref{dbscan}. After clustering, we follow the same operations in \cite{b9, b35} to merge very close eating episodes and remove very short episodes. Specifically, if the distance between two eating episodes is less than 3 min, they are merged into one episode. Additionally, if the duration of the episode is less than 3 min after merging, the eating episode is removed. Finally, we obtain a set of detected eating episode $\mathcal{M}=\left\{m_1, m_2, ...m_Z\right\}$, where each $m_j$ corresponds to the interval $[t_{j}^{l},t_{j}^{r}]$ ($1\leq{j}\leq{Z}$) that corresponds the left and right boundaries of $j$-th eating episode, $Z$ denotes the total number of detected eating episodes.

\vspace{-0.25cm}
\subsection{Eating Speed Estimation}
Eating speeds are obtained by utilizing the detected bite set $\mathcal{B}$ and eating episode set $\mathcal{M}$. The eating episode detection algorithm predicts the start point $t_{j}^{l}$ and end point $t_{j}^{r}$ of the $j$-th eating episode. If the detected bite $b_i$ falls within the interval of the eating episode, this bite is considered to belong to that eating episode. At the end, the number of bites divided by the duration of the eating episode gives the eating speed.

  \begin{figure}[t]
  \centering
  \includegraphics[scale=0.45]{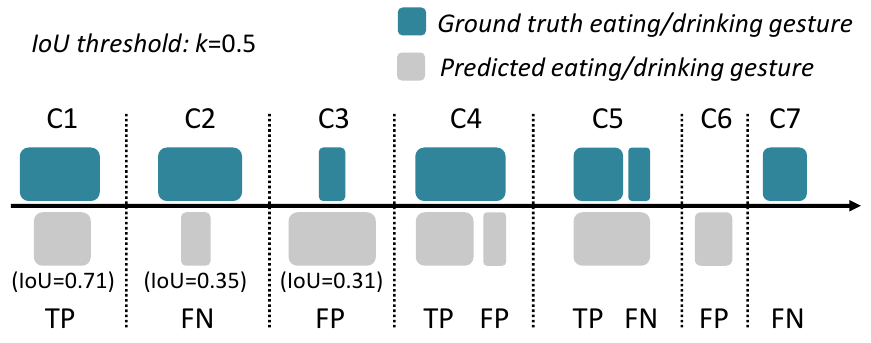}
  \vspace{-0.3cm}
  \caption{Segment-wise evaluation examples. If the target is drinking class, eating gestures are categorized as other, and vice versa.} 
  \label{seg_eva}
  \vspace{-0.3cm}
\end{figure}

\section{Evaluation and Experiment}
\subsection{Evaluation Criteria}
\subsubsection{Evaluation on Bite Detection}

The output of the proposed seq2seq model is point-wise multi-class prediction. As bite-related datasets are normally unbalanced, we choose to use the index Cohen Kappa \cite{b52} to represent the performance of point-wise classification. Although such results can indicate the performance of the model, it should be noted that the purpose of bite detection is to count the number of bites, whereas point-wise results are unable to reveal such information. To address this issue, we use a segment-wise evaluation method to evaluate the bite detection, which has been applied in previous study \cite{b26}. Fig. \ref{seg_eva} shows examples of this evaluation. The evaluation method involves two steps. Firstly, the intersection over union (IoU) between each predicted bite and ground truth bite is calculated, as shown in Fig. \ref{seg_eva} C1-C3. Secondly, the calculated IoU is compared to a selected threshold $k$ to determine segment-wise true positive (TP), false negative (FN) and false positive (FP). Subsequently, the segmental F1-score is calculated for each class (eating and drinking). The segment-wise evaluation allows for short temporal shifts between ground truth and prediction, which maybe caused by annotation variability. Meanwhile, it furnishes straightforward information including the number of detected bites. Furthermore, by adjusting the threshold $k$, we evaluate not only the detection performance, but also segmentation performance. Two thresholds are selected as 0.1 and 0.5.

\subsubsection{Evaluation on Eating Episode Detection}
The evaluation of eating episode detection mainly focuses on two aspects, specifically, the detection performance, and the segmentation performance (how well the boundaries of each eating episode are determined). Therefore, the aforementioned segment-wise evaluation method is also used for eating episode detection ($k=0.5$). Additionally, for each predicted eating episode, we utilize the IoU score to evaluate the segmentation performance.

\subsubsection{Evaluation on Eating Speed Estimation}
The mean absolute percentage error (MAPE) is used to evaluate the deviation between the estimated speed and the ground truth speed.
\begin{equation}\rm MAPE =\frac{1}{z}\sum_{i=1}^{z}{\left|\frac{\hat{s}_{i}-s_{i}}{s_{i}}\right|}\times 100\%\label{eq3}\end{equation}
where $z$ is the total number of truly detected meals (TP), $\hat{s}_{i}$ and $s_{i}$ represent the estimated eating speed and ground truth eating speed, respectively.

For statistical quantitative analysis, the Pearson correlation coefficient (PCC) is also calculated to assess the correlation of the predicted eating speed with the ground truth objectively.

\begin{equation}\rm PCC = \frac{{\sum_{i=1}^{z} (s_i - \bar{s})(\hat{s}_i - \bar{\hat{s}})}}{{\sqrt{{\sum_{i=1}^{z} (s_i - \bar{s})^2} \cdot \sum_{i=1}^{z} (\hat{s}_i - \bar{\hat{s}})^2}}}
\label{eq4}\end{equation}
where $\bar{s}$ and $\bar{\hat{s}}$ represent the mean value of the ground truth speed and estimated speed, respectively. It should be noted that the MAPE and PCC are only calculated among eating speeds of successfully detected eating episodes.

\vspace{-0.25cm}
\subsection{Models for Benchmarking}
To evaluate the efficacy of the proposed method, existing models for bite detection were chosen as comparative benchmarks. Specifically, the CNN-LSTM \cite{b9}, ResNet-LSTM \cite{b17}, and MS-TCN \cite{b43} were chosen. Additionally, the bidirectional-LSTM layer (BiLSTM) was used to replace the LSTM layer to compose the CNN-BiLSTM and ResNet-BiLSTM models as extra models. It should be noted that the output of models are point-wise probabilities. 

\begin{table}[b]
\vspace{-0.3cm}
\caption{Bite Detection Performance on FD-I Dataset}
\label{bite_results_2}
\begin{center}
\scalebox{0.60}{
\begin{tabular}{l|l|c|cc|cc}
\toprule
\hline
\multirow{2}*{Data}&\multirow{2}*{Model}&{Point-wise}&\multicolumn{2}{|c|}{Segmental Eating F1-score}&\multicolumn{2}{|c}{Segmental Drinking F1-score}\\
~&~&Kappa&{$k=0.1$}&{$k=0.5$}&{$k=0.1$}&{$k=0.5$}\\
\midrule

\multirow{6}*{FD-I}&CNN-LSTM& 0.557 & 0.738 & 0.583 & 0.811 & 0.619 \\
~&CNN-BiLSTM & 0.666& 0.788 & 0.694 & 0.859 & 0.762  \\
~&ResNet-LSTM& 0.630& 0.753 & 0.626 & 0.791 & 0.682 \\
~&ResNet-BiLSTM& 0.704& 0.790 & 0.701 & 0.849 & 0.779 \\
~&MS-TCN& 0.702& 0.824 & 0.761 & \textbf{0.906} & 0.853  \\
~&TCN-MHA & \textbf{0.735}&\textbf{0.849} & \textbf{0.781} & 0.902 & \textbf{0.858}   \\
\hline
\bottomrule
\end{tabular}}
\end{center}
\vspace{-0.4cm}
\end{table}

\begin{table}[b]
\vspace{-0.2cm}
\caption{Eating Episode and Speed Detection on FD-I Dataset}
\label{meal_results_1}
\begin{center}
\scalebox{0.65}{
\begin{tabular}{l|l|lllll|cc}
\toprule
\hline
\multirow{2}*{Data}&\multirow{2}*{Model}&\multicolumn{5}{|c|}{Eating Episodes Detection}&\multicolumn{2}{|c}{Eating Speed}\\
~&~&TP &FP&FN&F1&IoU&MAPE&PCC\\

\midrule
\multirow{6}*{FD-I}&CNN-LSTM&60 & 0 & 14 & 0.896 & \textbf{0.900} & 0.238 & 0.696 \\
~&CNN-BiLSTM&63 & 1 & 11 & 0.913 & 0.895 &  0.202 & 0.789  \\
~&ResNet-LSTM&65 & 6 & 9 & 0.897 & 0.863 &  0.197 & 0.782  \\
~&ResNet-BiLSTM&64 & 6 & 10 & 0.889 & 0.864 & 0.155 & 0.829 \\
~&MS-TCN &62 & 0 & 12 & 0.912 & 0.881  & 0.151 & 0.827 \\
~&TCN-MHA&64 & 0 & 10 & \textbf{0.928} & 0.899& \textbf{0.110} & \textbf{0.925} \\  
\hline
\bottomrule
\end{tabular}}
\end{center}
\vspace{-0.3cm}
\end{table}

\begin{figure*}[t]
      \centering
      \includegraphics[scale=0.5]{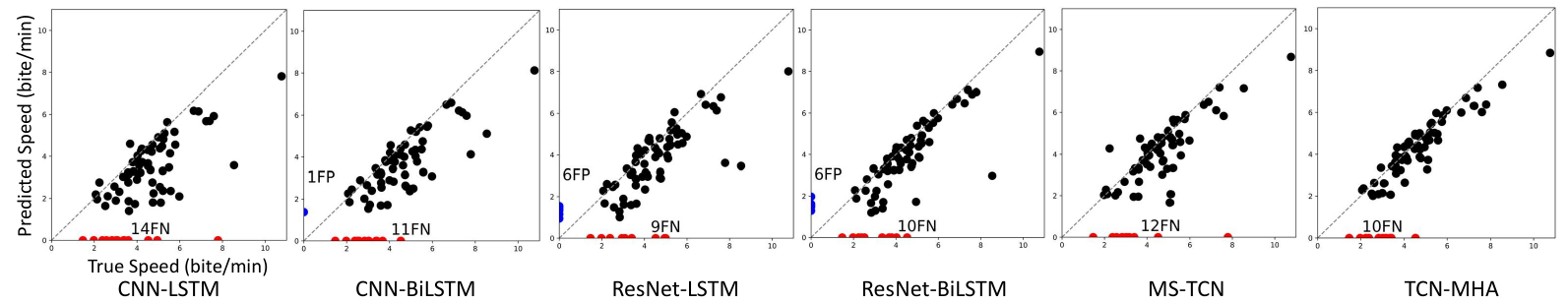}      
      \vspace{-0.2cm}
      \caption{The scatter plot for eating speed on FD-I dataset. Black dots represent TP eating episodes, red dots are FNs, blue dots are FPs.}   
      \label{speed_scatter_1}
\end{figure*}

\begin{figure*}[t]
      \centering
      \includegraphics[scale=0.27]{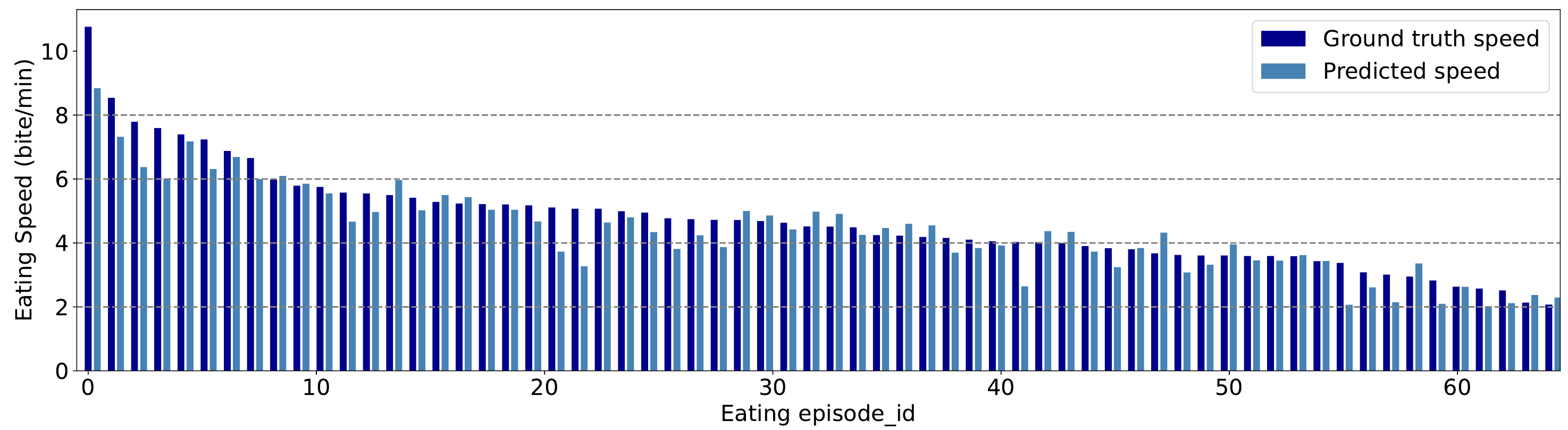}     
      \vspace{-0.2cm}
      \caption{The bar plot for eating speed on FD-I dataset.}   
      \vspace{-0.3cm}
      \label{speed_bar_plot_1}
\end{figure*}

\begin{table*}[t]
\vspace{-0.2cm}
\caption{In-meal Bite Detection, Eating Episode Detection and Eating Speed Performance on FD-II Dataset}
\label{meal_results_2}
\begin{center}
\scalebox{0.80}{
\begin{threeparttable} 
\begin{tabular}{c|l|cc|lllll|cc|c|c}
\toprule
\hline
\multirow{2}*{Data}&\multirow{2}*{Model}&\multicolumn{2}{|c|}{Eating Gesture F1-score$^a$}&\multicolumn{5}{|c|}{Eating Episodes Detection}&\multicolumn{2}{|c|}{Eating Speed}&\multirow{2}*{\# Paras (M)}&\multirow{2}*{\# FLOPs (G)}\\
~&~&{$k=0.1$}&{$k=0.5$}&TP &FP&FN&F1&IoU&MAPE&PCC&~&~\\
\midrule
\multirow{6}*{FD-II}&CNN-LSTM& 0.731& 0.555&43 & 53 & 9 & 0.581 & 0.746 & 0.231 & 0.683 &0.134  & 0.129  \\
~&CNN-BiLSTM&0.783 &0.622 &47 & 35 & 5 & 0.701 & 0.786 & 0.173 & 0.860  &0.241  & 0.233   \\
~&ResNet-LSTM&0.752 &0.584 &48 & 42 & 4 & 0.676 & 0.809 & 0.201 & 0.780 &3.078  & 2.955  \\
~&ResNet-BiLSTM&0.798 &0.650 &49 & 26 & 3 & 0.772 & 0.827 & \textbf{0.142} & 0.910  &3.415  & 3.280  \\
(Holdout)& MS-TCN &0.814 & 0.636&47 & 22 & 5 & 0.777 & 0.831 & 0.155 & 0.886 &0.298  & 0.284\\
~&TCN-MHA& \textbf{0.820}& \textbf{0.651}&48 & 7 & 4 & \textbf{0.897} & \textbf{0.841} & 0.146 & \textbf{0.924} &0.203 & 0.194\\ 
\hline
\bottomrule
\end{tabular}
    \begin{tablenotes}    
    \footnotesize              
    \item[a] The eating segmental F1-scores in FD-II dataset only show the results of bite detection in meal sessions. Eating gestures from the outside of meals are not labelled, hence is unable to be evaluated.        %
    \end{tablenotes}   
\end{threeparttable} 
}
\end{center}
\vspace{-0.3cm}
\end{table*}

\begin{table}[t]
\vspace{-0.3cm}
\caption{Results on OREBA and Clemson datasets using TCN-MHA}
\label{meal_results_3}
\begin{center}
\scalebox{0.68}{
\begin{threeparttable} 
\begin{tabular}{c|c|cc|ccccc|cc}
\toprule
\hline
\multirow{2}*{Valid}&\multirow{2}*{Data}&\multicolumn{2}{|c|}{Eating Gesture F1-score}&\multicolumn{5}{|c|}{Eating Episodes Detection}&\multicolumn{2}{|c}{Eating Speed}\\
~&~&{$k=0.1$}&{$k=0.5$}&TP &FP&FN&F1&IoU&MAPE&PCC\\
\midrule
\multirow{2}*{7-fold}&OREBA     & 0.897 & 0.878 & 97  & 0 & 4  & 0.980 & 0.967 & 0.069 & 0.940  \\
~&Clemson   & 0.894 & 0.835 & 445 & 2 & 12 & 0.985 & 0.931 & 0.097 & 0.913  \\
\midrule
\multirow{2}*{hold}&OREBA     & 0.803 & 0.720 & 88  & 0 & 13  & 0.931 & 0.907 & 0.119 & 0.834  \\
out&Clemson   & 0.710 & 0.652 & 275 & 0 & 182 & 0.751 & 0.825 & 0.264 & 0.607  \\
\hline
\bottomrule
\end{tabular}
\end{threeparttable} }
\end{center}
\vspace{-0.4cm}
\end{table}

\begin{table}[t]
\vspace{-0.2cm}
\caption{Eating styles and locations performance using TCN-MHA on FD-I dataset}
\label{type_results}
\begin{center}
\scalebox{0.68}{
\begin{threeparttable} 
\begin{tabular}{l|l|cccc|cc}
\toprule
\hline
\multirow{2}*{View}&\multirow{2}*{Type}&\multicolumn{4}{|c|}{Eating Episodes Detection}&\multicolumn{2}{|c}{Eating Speed}\\
~&~&TP&FN&Recall&IoU&MAPE&PCC\\
\midrule
\multirow{4}*{Cutlery}&Fork\&knife  & 34 &  1 & 0.971 &0.942  & 0.099 & 0.866  \\
~&Chopsticks                        & 6  &  1 & 0.857 &0.961  & 0.136 & 0.982  \\
~&Spoon                             & 8  &  1 & 0.889 &0.955  & 0.079 & 0.921  \\
~&Hand                              & 16 &  7 & 0.696 &0.896  & 0.139 & 0.940  \\
\midrule
\multirow{3}*{Location}&Home        & 12 & 4 & 0.750 &0.922 & 0.100 & 0.956  \\
~&Restaurant                        & 31 & 0 & 1.000 &0.943 & 0.113 & 0.923  \\
~&Workspace                        & 21 & 6 & 0.778 &0.927 & 0.111 & 0.924  \\
\hline
\bottomrule
\end{tabular}
\end{threeparttable} }
\end{center}
\vspace{-0.5cm}
\end{table}

\section{Experimental Results}

\subsection{Experiments on FD-I Dataset}
\subsubsection{Bite Detection}
The 7-fold cross validation was used on FD-I dataset. Specifically, within each fold, all corresponding IMU data from participants designated in the test list were allocated to the test set to avoid information leakage. As mentioned in Section \uppercase\expandafter{\romannumeral3}-C, the FD-I dataset is highly unbalanced, hence, the MO and OREBA datasets were included as part of the training set. Results are shown in Table \ref{bite_results_2}. For point-wise results, TCN-MHA obtained the highest kappa score (0.735). For segment-wise results, the TCN-MHA also achieved the highest F1-score for eating with the value of 0.849 and 0.781, for $k=0.1$ and $k=0.5$, respectively. However, MS-TCN yielded a higher F1-score for drinking compared to TCN-MHA (0.906$\rightarrow$0.902, for $k=0.1$).

\subsubsection{Eating Episode Detection}
The predicted bite sequences from previous step were clustered into eating episodes using the DBSCAN-based algorithm. The results of eating episodes detection are shown in Table \ref{meal_results_1}. Among the 74 ground truth eating episodes, the TCN-MHA successfully detected 64 sessions with a mean IoU of 0.899.

\subsubsection{Eating Speed}
Table \ref{meal_results_1} shows the eating speed estimation results. The TCN-MHA model had the least MAPE of 0.110; the highest PCC value of 0.925. The scatter plots are drawn to visualize the correlation between predicted and ground truth eating speed in the FD-I dataset (Fig. \ref{speed_scatter_1}). Most eating episodes had an eating speed falling into the range of 2-6 bite/min. Additionally, the result for each individual episode on TCN-MHA model has been shown in Fig. \ref{speed_bar_plot_1}.

\begin{table*}[t]
\caption{Existing studies on Automated Food Intake Monitoring in full-day scenarios}
\label{compare_litera}
\begin{center}
\scalebox{0.85}{
    \begin{threeparttable} 
    \begin{tabular}{l|l|l|l|l|l|c|c|c}
    \toprule
    \hline
        \multirow{2}*{Work} & \multirow{2}*{Position$^a$} & \multirow{2}*{Sensor$^b$} & \multirow{2}*{\# Participants} & \multirow{2}*{\# Days} & \multirow{2}*{\# Hours} &Eating Episode &Bite annotation & Eating \\ 
         ~ & ~ &  ~ & ~ & ~  & ~ & detection&\& detection$^c$ & speed\\
\midrule
        Dong \emph{et al}. (2014) \cite{b30}&  P1 & S1 & 43 & 43 & 449 & \checkmark& - & - \\ 
        Fontana \emph{et al}. (2014) \cite{b31}&  P1, P2, P3 & S2, S3, S4 & 12 & 12  & -& \checkmark & - & - \\ 
        Thomaz \emph{et al}. (2015) \cite{b32}&  P1 & S1 & 8 & 37  & - & \checkmark& - & - \\ 
        Mirtchouk \emph{et al}. (2017) \cite{b33}  &  P1, P2, P4 & S1, S5 & 11 & 25  & 257& \checkmark & - & - \\ 
        Bedri \emph{et al}. (2017) \cite{b35}&  P2, P5 & S1, S5, S6 & 10  & - & 45& \checkmark & - & - \\ 
        Sen \emph{et al}. (2018) \cite{b36}& P1 & S1, S7 & 9  & - & 52 & \checkmark& - & -  \\ 
        Schiboni \emph{et al}. (2018) \cite{b38}& P1 & S1 & 7  & 35 & 345 & - & \checkmark & -  \\
        Sharma \emph{et al}. (2020) \cite{b13,b37} &  P1 & S1 & 351 & 351  & 4,068& \checkmark & - & - \\ 
        Doulah \emph{et al}. (2020) \cite{b14}&  P4 & S2, S7, S8 & 30 & 60  & - & \checkmark& - & - \\ 
        Zhang \emph{et al}. (2020) \cite{b19} & P5 & S1, S6, S9 & 20 & -  & 271 & \checkmark & - & -\\
        Bedri \emph{et al}. (2020) \cite{b18} & P4 & S1, S6, S7 & 23 & 8  & 91& \checkmark & \checkmark & -\\ 
        Kyrtisis \emph{et al}. (2020) \cite{b9}  & P1 & S1 & 12 & 12  & 113& \checkmark & - & - \\ 
        \textbf{Ours} &  \textbf{P1} & \textbf{S1} & \textbf{61} & \textbf{61} & \textbf{513} & \checkmark & \checkmark & \checkmark \\ 
\hline
\bottomrule
\end{tabular}
    \begin{tablenotes}    
    \footnotesize              
    \item[a] P1: Wrist, P2: Ear, P3: Chest, P4: Head, P5: Neck. %
    \item[b] S1: IMU, S2: Accelerometer, S3: Piezo, S4: RF, S5: Microphones, S6: Proximity, S7: Camera, S8: Flex, S9: Light. %
    \item[c] It should be noted that some approaches use in-meal datasets to train the model to detect bite, then use the trained model to process free-living datasets, but these free-living datasets do not contain bite-level label, so they are considered no bite detection \& evaluation in free-living scenarios.%
    \end{tablenotes}   
\end{threeparttable} 
}
\end{center}
\vspace{-0.3cm}
\end{table*}

\vspace{-0.25cm}
\subsection{Experiments on FD-II Dataset}
To further validate the proposed method on eating speed measurement, the FD-II was used as the holdout dataset. We utilized two in-meal datasets (OREBA, MO) and the FD-I dataset to train our model, then used the entire FD-II as the test set. It should be noted that we were only able to measure the eating speed during meal sessions, as we lack labels for out-of-meal sessions. Hence, we did not evaluate the predicted snack sessions (eating episodes with duration less than 7 min). The performance of in-meal bite detection, meal detection, and speed measurement are shown in Table \ref{meal_results_2}. For in-meal bite detection, the TCN-MHA model yielded the highest segmental F1-score with 0.820 when $k=0.1$. The TCN-MHA obtained the best performance in meal detection, which successfully detected 48 meal sessions (7 FPs, 4 FNs), with an F1-score of 0.897 and a mean IoU of 0.841. For eating speed, the TCN-MHA had the MAPE of 0.146 and PCC of 0.924. The ResNet-BiLSTM had the highest number of TP meals, however the number of FPs was also higher than that of TCN-MHA.

\vspace{-0.25cm}
\subsection{Experiments on Public Datasets}
To compare the performance across datasets, the public OREBA and Clemson datasets were used as benchmarks. Both 7-fold and holdout validation methods were employed. In the 7-fold validation, the split of test set in each fold was participant-level to avoid information leakage. The TCN-MHA model was trained and evaluated within each dataset separately. For holdout validation, the model was trained on FD-I and MO datasets and then tested on OREBA and Clemson, respectively. Results are shown in Table \ref{meal_results_3}. In 7-fold validation, it is evident that higher F1-scores for eating gesture detection and episodes detection were obtained on the OREBA and Clemson datasets compared to that of FD-I dataset (Table \ref{bite_results_2} and Table \ref{meal_results_1}). The OREBA achieves the lowest MAPE of 0.069 with the PCC of all datasets exceeding 0.900. These results validate the performance of proposed approach, implying that it is indeed more challenging to detect eating gestures and episodes in near-free-living environments. Results also show that measuring eating speed in an uncontrolled environment (i.e., cafeteria) is more challenging than in a controlled environment (i.e., lab). In holdout validation, all evaluation metrics decreased compared to those in the 7-fold validation. On OREBA dataset, the MAPE of eating speed increased from 0.069 to 0.119. The reduction on the Clemson was much larger than that on OREBA dataset, the MAPE on Clemson increased from 0.097 to 0.264.

\begin{table}[t]
\vspace{-0.2cm}
\caption{Dominant hand performance on FD-I dataset using TCN-MHA}
\label{dom_results}
\begin{center}
\scalebox{0.7}{
\begin{threeparttable} 
\begin{tabular}{l|cc|lllll|cc}
\toprule
\hline
\multirow{2}*{Hands}&\multicolumn{2}{|c|}{Eating Gesture F1-score}&\multicolumn{5}{|c|}{Eating Episodes Detection}&\multicolumn{2}{|c}{Eating Speed}\\
~&{$k=0.1$}&{$k=0.5$}&TP &FP&FN&F1&IoU&MAPE&PCC\\
\midrule
Both hands& 0.849& 0.781&64 & 0 & 10 & 0.928 & 0.899 & 0.110 & 0.925  \\
Dominant hand& 0.806 &0.738 & 56 & 0 & 18 & 0.862 & 0.857 & 0.128 & 0.854  \\
\hline
\bottomrule
\end{tabular}
\end{threeparttable} 
}
\end{center}
\vspace{-0.5cm}
\end{table}

\section{Discussion}
In this study, we first tested several models on FD-I and FD-II for bite detection. Results from Table \ref{bite_results_2} show that bite detection in near-free-living environments is feasible. Meanwhile, Table \ref{meal_results_1} also indicates a tendency for models to overlook certain eating episodes, as evidenced by a higher number of FNs compared to FPs. Upon comparing the output and the annotation video, we found that FNs are mainly from snack sessions, implying that the clustering of snack sessions is more challenging than detecting meals \cite{b13,b33}. One reason is that the eating patterns of snacking can vary widely in terms of frequency and duration, making it difficult to cluster all snacking gestures into specific episodes. 

A deeper analysis has also been conducted to compare the model’s performance across different eating styles and environments, as shown in Table \ref{type_results}. Regarding four eating styles, it is discerned that eating with hand exhibits the lowest performance in terms of both Recall and IoU scores in episodes detection, as well as the MAPE in eating speed measurement step. This implies that the model tends to miss hand eating episodes. Based on observation, we found that eating with hand is more difficult to detect. This difficulty is attributed to the inherent variability associated with this eating style. Additionally, hand-eating gestures closely resemble other hand movements commonly encountered in daily life, further complicating the detection. The challenge of detecting snacking sessions is also linked to such issue. Among three types of eating locations, eating in restaurants has shown the highest performance by detecting all episodes and achieving the highest IoU score. One potential reason is that individuals tend to eat in a more formalized manner in restaurant, often using utensils such as fork \& knife. By checking the data, the FNs in workspace are mainly snacking sessions.

\begin{figure*}[t]
      \centering
      \includegraphics[scale=0.55]{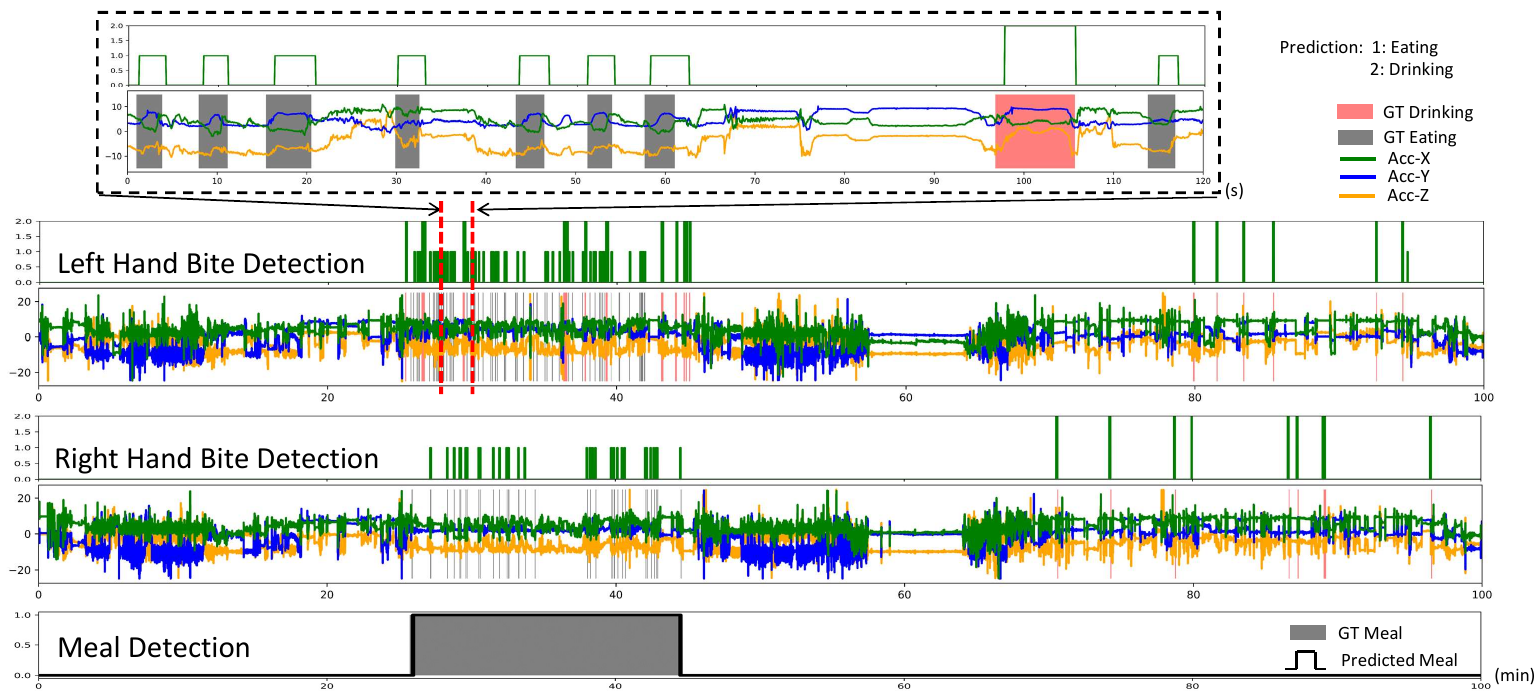} 
      \vspace{-0.2cm}
      \caption{The 100 min IMU data segment from both hands and the corresponding bite and meal detection examples. Additionally, a 2-min in-meal segment is also selected to show the detailed bite detection.} 
      \vspace{-0.3cm}
      \label{data_example}
\end{figure*}

When comparing our work to prior studies in full-day scenarios (as shown in Table \ref{compare_litera}), the dataset exhibits a comparable size to others, except for the dataset used by Sharma et al. \cite{b13,b37}, which is substantially larger than the rest. However, it's worth noting that existing free-living datasets mainly focus on eating episode detection, which only requires labeling the starting and ending times of eating episodes. In contrast, our data for eating speed measurement requires bite-level annotation, which demands additional efforts for data collection and annotation. The detection granularity of this study is shown in Fig. \ref{data_example}. Compared to eyeglass-based and necklace based approaches, one limitation of wristband-based approach is that we need to wear IMUs on both hands. Wearing the IMU only on the dominant hand may lead to the omission of some eating gestures, as illustrated in Table \ref{dom_results}. In comparison to utilizing IMU data from both hands, \#FNs increased from 10 to 18 due to the absence of bites from non-dominant hand, the MAPE increased from 0.110 to 0.128. These results suggest that using IMU data from only the dominant hand leads to missed detections of eating episodes. Additionally, for the eating speed from detected episodes, the MAPE performance also reduced. However, considering that people typically wear only one watch in normal daily life, the performance may still be acceptable.

Several wrist-worn IMU based bite detection datasets have been published, however, they are only used separately to benchmark performance of different models. In this study, we examined the feasibility of combining different datasets for training. When performing bite detection on the FD-I dataset, our own MO dataset and the external public OREBA dataset were included as part of training set. This is the first attempt to integrate different datasets for food intake monitoring, resulting in a 0.7\% increase of F1-score in bite detection, 3.4\% increase of PCC in eating speed detection. The 7-fold validation results on Clemson and OREBA datasets showed in Table \ref{meal_results_3} demonstrate that the proposed approach performs better within meal-session datasets. Whereas the holdout validation results exhibit lower performance, especially for the Clemson dataset. One potential reason could be attributed to the orientation of the IMU sensor used in Clemson, which differs from that of the FD-I and OREBA. This suggests that the generalization ability of proposed approach may be limited by the orientation of IMU. Table \ref{meal_results_3} also shows that the largest dataset (Clemson) exhibits the highest variability among meal-session datasets, in terms of number of subjects, types of foods, and types of behaviors. This variability is also illustrated in the ground truth plots of eating speed in Fig. \ref{speed_violin}.

Regarding the scatter plot in Fig. \ref{speed_scatter_1}, we noticed a common outlier point in CNN-(Bi)LSTM and ResNet-(Bi)LSTM models, representing an eating episode with a ground truth speed of 8.54 bites/min. For ResNet-BiLSTM, the predicted speed is 2.98 bites/min. By inspecting the video, we found this outlier episode involved eating snacks with hand. This observation aligns with findings from eating style analysis, suggesting that hand-eating gestures are more challenging to detect.

\begin{figure*}[t]
      \centering
      \includegraphics[scale=0.33]{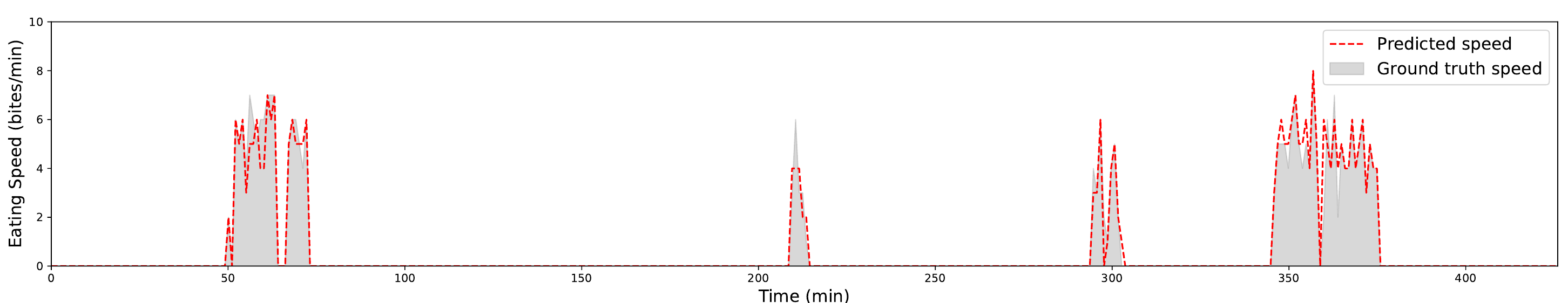}   
      \vspace{-0.2cm}
      \caption{The minute-level eating speed distribution in a full day from one participant in FD-I dataset.} 
      \vspace{-0.3cm}
      \label{minute_level}
\end{figure*}

To assess the model’s complexity, the number of parameters for each model and their floating point operations (FLOPs) are indicated in Table \ref{meal_results_2}. The number of parameters in ResNet-(Bi)LSTM-based models were significantly larger than CNN-LSTM (0.134 M) and TCN-MHA models (0.203 M). Additionally, a test was carried out to assess the latency for processing 1 min data using the TCN-MHA model. Utilizing a laptop equipped with an Intel Core i7 10750 CPU @2.6 GB, 6 cores (no GPU configuration), the TCN-MHA required 37.61 ms to generate predictions.

\begin{table}[t]
\caption{Minute-level Eating Speed}
\label{minute_results_1}
\begin{center}
\scalebox{0.7}{
\begin{tabular}{l|cc|cc}
\toprule
\hline
\multirow{2}*{Model}&\multicolumn{2}{|c|}{FD-I}&\multicolumn{2}{|c}{FD-II}\\
~&MAPE&PCC&MAPE&PCC\\
\midrule
CNN-LSTM& 0.256 & 0.702 & 0.304 & 0.650 \\
CNN-BiLSTM & 0.221 & 0.781 & 0.272 & 0.758\\
ResNet-LSTM& 0.246 & 0.747 & 0.326 & 0.710\\
ResNet-BiLSTM& 0.219 & 0.793& 0.273 & 0.797 \\
MS-TCN& 0.184 & 0.805& 0.224 & 0.816 \\
TCN-MHA & \textbf{0.181} & \textbf{0.840}& \textbf{0.212} & \textbf{0.834} \\
\hline
\bottomrule
\end{tabular}}
\end{center}
\vspace{-0.5cm}
\end{table}

The proposed wrist-worn IMU-based approach has the potential to replace the questionnaire-based surveys on eating speed in nutrition studies, advancing the analysis of the correlation between eating speed and obesity-related problems. Moreover, such a dietary-related digital biomarker can be applied to individuals interested in documenting their daily dietary habits. Further, combining data from fitness trackers, dietary analysis apps, and physiological sensors can generate a more comprehensive picture of an individual's health status, facilitating a holistic understanding of individual health profiles. For diabetes management, such technology can be integrated with the wearable continuous glucose monitor (CGM) sensors to gain deeper insights into postprandial glucose fluctuations and their correlation with eating behaviors. By investigating correlations between eating speed, meal composition, and postprandial glucose levels, predictive algorithms could be developed to anticipate glycemic responses to specific meals, thus empowering individuals with diabetes to proactively manage their blood glucose levels and mitigate the risk of hyperglycemia or hypoglycemia.

This study mainly focus on eating speed measurement for each eating episode, resulting in an averaged speed per episode, which is a well-accepted definition. Additionally, we also explored the feasibility of measuring minute-level eating speed. To achieve this, the number of detected bites in each minute is considered as minute-level eating speed. Fig. \ref{minute_level} shows the minute-level eating speed distribution through one day. Results are shown in Table \ref{minute_results_1}. The TCN-MHA had the MAPE of 0.181 and the PCC of 0.840 on FD-I, and had the MAPE of 0.212 and the PCC of 0.834 on FD-II. The meal-level eating speed represents a holistic view of an eating session, whereas minute-level speed provides insights into eating patterns at a more granular level and is suitable for studying immediate speed changes during a meal. On the other hand, the performance of minute-level speed detection (Table \ref{minute_results_1}) implies that the minute-level eating speed detection is more challenging compared to meal-level speed detection.

The collected dataset served to validate the feasibility of eating speed measurement in near-free-living environments. Participants were recorded in their habitual environments where they usually eat, work, and study, thereby ensuring a sense of familiarity with their surroundings. Nevertheless, one limitation is that the participants behaviors may potentially be affected when they were followed and recorded all day, which is the reason that we use the term \emph{near-free-living environments}. Additionally, the dataset only contains conventional eating behaviors, namely structured meal consumption and inter-meal snacking. Other types of eating behavior such as grazing, eating while walking, and night eating, were not included in the study. A larger dataset with a broader spectrum of eating behaviors would be helpful to further validate the robustness of the approach in the future. Meanwhile, the proposed method were based on off-line processing. It is worthwhile to exploit the feasibility of implementing this method to smartwatch-smartphone setup for daily eating speed monitoring. Another limitation is that obtaining fine-annotated ground truth data is troublesome. In our approach, research assistants had to follow participants activities to record videos. Existing wearable-based cameras \cite{b6} can be used for recording to minimize the effort in future study. Furthermore, to quantify the actual food intake, the wearable IMU system has the potential to be combined with the smart plate \cite{b47} and the smart snack box \cite{b51} to estimate the calorie intake in real life.

\section{Conclusion}
In this work, we presented a comprehensive framework for automated measurement of eating speed in near-free-living environments. To the best of our knowledge, this is the first of its kind. This framework has the potential to extend the application scope of the automated food intake monitoring field. The proposed method is built upon two essential tasks in automated food intake monitoring domain: bite detection and eating episode detection. The success of bite detection paves the way for eating episode detection and segmentation, resulting in a good capability for eating speed measurement. The holdout experiments further underscore its robustness in meal-level eating speed detection.

\section*{Acknowledgment}
The authors thank the participants involved in the experiments for their dedicated contributions of time and effort.

\end{document}